\documentclass[%
 aip,
cp,  
 amsmath,amssymb,
 reprint,%
]{revtex4-2}

\usepackage{graphicx}
\usepackage{dcolumn}
\usepackage{bm}

\usepackage[utf8]{inputenc}
\usepackage[T1]{fontenc}
\usepackage{mathptmx} 

\newtheorem{thm}{Theorem}

\newtheorem{prop}[thm]{Proposition}

\newtheorem{definition}[thm]{Definition}

\newtheorem{remark}[thm]{Remark}

\newcommand{\be}{\begin{equation*}}
\newcommand{\ee}{\end{equation*}}
\newcommand{\ben}{\begin{equation}}
\newcommand{\een}{\end{equation}}
\newcommand{\beqa}{\begin{eqnarray*}}
\newcommand{\eeqa}{\end{eqnarray*}}
\newcommand{\beqan}{\begin{eqnarray}}
\newcommand{\eeqan}{\end{eqnarray}}
\newcommand{\nn}{\nonumber}

\def\C{\mathbb{C}}
\def\R{\mathbb{R}}

\def\cK{\mathcal{K}}
\def\Sym{\mathrm{Sym}}

\newcommand{\pd}{\partial}
\def\dd{\mathrm{d}}

\newcommand{\Tr}{\mathrm{Tr}}

\def\cC{\mathcal{C}}

\def\cG{\mathcal{G}}

\def\cL{\mathcal{L}}
\def\cM{\mathcal{M}}
\def\cN{\mathcal{N}}
\def\cP{\mathcal{P}}
\def\cS{\mathcal{S}}

\def\cX{\mathcal{X}}

\def\bLambda{\boldsymbol{\Lambda}}

\def\mD{\mathbb{D}}

\def\rD{\mathrm{D}}

\def\vol{\mathrm{vol}}

\def\bm{\mathbf{m}}

\newcommand{\eqdef}{\stackrel{{\rm def.}}{=}}

\def\grad{\mathrm{grad}}
\def\Hess{\mathrm{Hess}}

\def\Re{\mathrm{Re}}
\def\Im{\mathrm{Im}}

\begin{document}

\title{Noether Symmetries of Two-Field Cosmological Models}

 \author{Lilia Anguelova}
 \email{anguelova@inrne.bas.bg }
\affiliation{
  Institute for Nuclear Research and Nuclear
  Energy, BAS, 1784 Sofia, Bulgaria
}
\author{Elena Mirela Babalic} 
 \email[Corresponding author: ]{mbabalic@theory.nipne.ro}
\affiliation{Department of Theoretical Physics, Horia Hulubei
National Institute for Physics and Nuclear Engineering, \\
Bucharest - Magurele 077125, Romania.}

\author{Calin Iuliu Lazaroiu}%
 \email{calin@ibs.re.kr.}
\affiliation{%
Center for Geometry and Physics, Institute for Basic Science, Pohang 37673, Republic of
  Korea
}%
\affiliation{Department of Theoretical Physics, Horia Hulubei
National Institute for Physics and Nuclear Engineering, \\
Bucharest - Magurele 077125, Romania.}

\date{\today} 

\begin{abstract}
We summarize our work on ``hidden'' Noether symmetries of multifield cosmological models and
the classification of those two-field cosmological models which admit such symmetries. 
\end{abstract}

\maketitle

\section{Introduction}

Cosmological models with at least two real scalar fields are of
increasing interest in cosmology (for example, as
models of quintessence as well as in descriptions of inflation which
may allow for contact with string theory).  Careful study shows that
certain such models possess ``hidden'' Noether symmetries
\cite{ABL1,ABL2}, which -- when present -- allow one to simplify
various problems and sometimes to construct exact solutions which are
not accessible through other means. We summarize our recent work aimed
at the classification of $n$-field cosmological models admitting such
symmetries. For the case of two real scalar fields, our results show that
all such models are of generalized $\alpha$-attractor type
\cite{genalpha} with scalar manifold a disk, a punctured disk or
an annulus \cite{elem}, endowed with a complete metric with fixed value $K=-3/8$
of the Gaussian curvature.

\section{\label{sec:level1}FLRW cosmology with scalar matter}

\begin{definition}
A {\bf scalar triple} is an ordered system $(\cM,\cG,V)$,
where:
\begin{itemize}
\item $(\cM,\cG)$ is a Riemannian $n$-manifold (called {\em scalar
  manifold})
\item $V\in \cC^\infty(\cM,\R)$ is a smooth real-valued function
  defined on $\cM$ (called {\em scalar potential}).
\end{itemize}
\end{definition}
\noindent We shall assume throughout the paper that $\cM$ is connected,
oriented and of dimension $n\ge2$ and that $\cG$ is complete.  Moreover, we assume for
simplicity that $V>0$ on $\cM$. In most applications to cosmology, the
scalar manifold is non-compact; however, notice that it need {\em not}
be simply-connected. 

Each scalar triple defines a {\bf scalar cosmological model}, which
describes Einstein gravity on a space-time with topology $\R^4$,
coupled to $n=\dim\cM$ scalar fields described by a smooth map
$\varphi\in\cC^\infty(\R^4,\cM)$ from the space-time to the scalar
manifold $\cM$; the scalar potential describes self-interaction of the
scalar fields.  The cosmological model has action:
\ben
\label{Action}
\cS_{\cM,\cG,V}[g,\varphi] = \int_{\R^4}  \left[ \frac{R(g)}{2} - \frac{1}{2}
  \Tr_g\varphi^\ast(\cG) - V \circ\varphi \right]~\vol_g ~,
\een
where the space-time metric $g$ has `mostly plus' signature. Locally, we have $\Tr_g\varphi^\ast(\cG)=g^{\mu\nu}\cG_{ij}\partial_\mu\varphi^i\partial_\nu\varphi^j$, 
where we use the Einstein summation convention. Completeness of the scalar manifold metric $\cG$ ensures
conservation of energy in such models. 

The metric $g$ of simply connected and spatially flat {\bf FLRW
  universe} $(\R^4,g)$ has the following squared line element in
global Cartesian coordinates $(t=x^0,x^1,x^2,x^3)$, where the strictly
positive smooth function $a\in \cC^\infty(\R,R_{>0})$ is called the
{\em scale factor}:
\ben
\label{FLRW}
\dd s_g^2 := - \dd t^2 + a(t)^2 \dd \vec{x}\,^2~~,~~\mathrm{where}~~\vec{x}=(x^1,x^2,x^3)~~.
\een
This implies $\vol_g=\sqrt{|\det g|}\,\dd^3\vec{x}\,\dd t=a(t)^3 \dd^3 \vec{x} \,\dd t$. 
Spatially homogeneous scalar field configurations in an FLRW universe are described by maps $\varphi$ which
depend only on the cosmological time $t$:
\ben
\label{PhiHom}
\varphi(t,\vec{x})=\varphi(t)~~.
\een
This gives $\Tr_g\varphi^\ast(\cG)=||\dot{\varphi}||^2_\cG$, where $\dot{~}\eqdef \frac{\dd}{\dd t}$~~.

\section{The minisuperspace Lagrangian and the Friedmann constraint}

\noindent Substituting \eqref{FLRW} and \eqref{PhiHom} in
\eqref{Action} and ignoring integration over
$\vec{x}$ produces the {\em minisuperspace action}:
\ben
\label{cS}
S_{\cM,\cG,V}[a,\varphi]=\int_{-\infty}^{\infty} \! \dd t \, L_{\cM,\cG,V}(a(t),\dot a(t),\varphi(t),\dot{\varphi}(t))~~,
\een
where the {\em minisuperspace Lagrangian} takes the form:
\ben 
\label{L}
L_{\cM,\cG,V}(a,\dot a,\varphi,\dot{\varphi})=
a^3\left[-3H^2+\frac{1}{2} ||\dot{\varphi}||^2_\cG - V\circ \varphi\right]~~.
\een
Here $H\eqdef\frac{\dot{a}}{a}$ is the Hubble parameter. This autonomous Lagrangian
describes a mechanical system with $n+1$ degrees of freedom and
configuration space $\cN\!=\! \R_{>0}\!\times \!\cM$.  The $g^{00}$
component of the Einstein equations gives the {\em Friedmann
  constraint}:
\ben
\label{Friedmann}
\frac{1}{2}||\dot{\varphi}||_\cG^2+V(\varphi)=3 H^2~,
\een
which is equivalent with the zero energy shell condition.
Indeed, the canonical momenta of $L$ are:
\ben
\label{pa}
p_a = \frac{\pd L}{\pd \dot{a}}=-6 a \dot{a}=-6 a^2 H~~,~~p_{i} = \frac{\pd L}{\pd \dot{\varphi}^i}=\cG_{ij}(\varphi) a^3 \dot{\varphi}^j=\cG_{ij}p^j~~
\een
while the phase space is the cotangent bundle $\cP\eqdef
T^\ast\cN\simeq \R_{>0}\times \R \times T^\ast \cM$ to the
configuration space, endowed with its canonical symplectic structure with local form:
\be
\omega=\dd p_a \wedge \dd a+\dd p_{i}\wedge \dd \varphi^i~~.
\ee
The {\em minisuperspace Hamiltonian} is:
\be
{\cal H}=p_a \dot{a}+p_{i}\dot{\varphi}^i-L
\!=\! -\frac{1}{12a} p_a^2 +\frac{1}{2a^3} ||p||^2_\cG + a^3 V\circ \varphi
\!=\! a^3 \left[-3 H^2 \!+\!\frac{1}{2} ||\dot{\varphi}||^2_\cG \!+\! V\circ \varphi\right]~~
\ee
and hence the Friedmann constraint amounts to requiring ${\cal H}=0$. Since the Hamiltonian 
is conserved, the Friedmann constraint \eqref{Friedmann} amounts to an algebraic relation between the integration constants of the solutions to the Euler-Lagrange equations of \eqref{L}. 
 
\subsection{The Euler-Lagrange and cosmological equations}

\noindent The Euler-Lagrange equations of \eqref{L} are equivalent with:
\beqan
\label{ELred}
&&3H^2+2\dot{H}+\frac{1}{2} ||\dot{\varphi}||^2_\cG - V(\varphi)=0\\
&& (\nabla_t +3H)\dot{\varphi} + (\grad_\cG V)(\varphi)=0~~ \nn
\eeqan
where:
\beqa
&&\nabla_t\dot\varphi^i= \ddot \varphi^i + \Gamma^i_{jk}\dot \varphi^j\dot \varphi^k \\
&&\grad_\cG V=(\grad_\cG V)^i\pd_i=\cG^{ij}\pd_j V\pd_i~~
\eeqa
and $\pd_j :=\frac{\pd }{\pd \varphi^j}$~.

\begin{prop}
When supplemented with the Friedmann constraint, the Euler-Lagrange equations \eqref{ELred}
are equivalent with the {\bf cosmological equations}:
\beqan
\label{eom}
6H^2 -||\dot{\varphi}||^2_\cG- 2V(\varphi) &=& 0~~\nn\\
\nabla_t \dot{\varphi}+3H \dot{\varphi}+(\grad_{\cG} V)( \varphi) &=&0~~ . 
\eeqan
\end{prop}

\begin{remark}
When $\dot{a}>0$, one can eliminate $H$ algebraically from the first
cosmological equation as: 
\be
H(t)=\frac{1}{\sqrt{6}}\sqrt{||\dot{\varphi}(t)||_\cG^2+2V(\varphi(t))}~~.
\ee
Then \eqref{eom} gives the {\em reduced cosmological equation} for $\varphi(t)$:
\ben
\label{eomsingle}
\nabla_t \dot{\varphi}(t)+\sqrt{\frac{3}{2}}
\sqrt{||\dot{\varphi}(t)||_\cG^2+2V(\varphi(t))}~\dot{\varphi}(t)+
(\grad_{\cG} V)(\varphi(t))=0~~,
\een
which defines a (dissipative) {geometric dynamical system} on $T\cM$.
\end{remark}

\section{Variational symmetries}
\noindent The tangent space to the configuration space
$\cN=\R_{>0}\times \cM$ has the decomposition $T\cN=T_{(1)}\cN \oplus
T_{(2)} \cN$, where $T_{(1)}$ and $T_{(2)}$ are the pullbacks of the vector bundles $T\R_{>0}$ and $T\cN$ through the canonical projections.  Accordingly,
any vector field $X\in \cX(\cN)$ decomposes as:
\be
X=X_{(1)}+X_{(2)}~~.
\ee
In local coordinates $(U,a,\varphi^i)$ on $\cN$, we have:
\be
{}{X_{(1)}(a,\varphi)=X^a(a,\varphi) \frac{\pd}{\pd a}~~,~~X_{(2)}(a,\varphi)=X^i(a,\varphi) \frac{\pd}{\pd \varphi^i}}~~,~~X=X^a(a,\varphi) \frac{\pd}{\pd a}+X^i(a,\varphi) \frac{\pd}{\pd \varphi^i}~~.
\ee

\begin{definition}
A (strong) variational symmetry of $L$ is a vector field $X\in \cX(\cN)$
which satisfies the {}{Noether symmetry} condition:
\ben
\label{NoetherSymCond}
\cL_{X^1} (L)=0~~.
\een
Here $X^1\in \cX(T\cN)$ is the first jet prolongation of $X$, with local form:
\vspace{-2mm}
\be
X^1=X(a,\varphi)+\dot X^a(a,\varphi)\frac{\pd}{\pd\dot a}+\dot X^i(a,\varphi)\frac{\pd}{\pd \dot \varphi^i}
\ee
\end{definition}

\noindent The Noether symmetry condition takes the local form:
\ben
\label{Noether}
X^a\frac{\pd L}{\pd a}+X^i\frac{\pd L}{\pd \varphi^i}
+\dot X^a\frac{\pd L}{\pd\dot a}+
\dot X^i\frac{\pd L}{\pd\dot \varphi^i}=0~~,
\een
where the total time derivative $\dot{\lambda}$ of a locally-defined function $\lambda$ on the configuration space
has the local expression: 
\be
\dot \lambda(a,\varphi):=\frac{\pd \lambda}{\pd t}+\frac{\pd \lambda}{\pd a} \dot a+\frac{\pd \lambda}{\pd \varphi^i} \dot \varphi^i~~.
\ee
\subsection{The characteristic system for variational symmetries}

\begin{thm}{\rm \cite{ABL2}}
\label{thm:Noether}
For the Lagrangian \eqref{L}, the Noether symmetry condition
 amounts to the requirement that $X_{(1)}$ and $X_{(2)}$ have the
following forms:
\be
\label{Xsol}
X_{(1)}(a,\varphi)=\frac{\Lambda(\varphi)}{\sqrt{a}}\frac{\pd}{\pd a}~~,~~X_{(2)}(a,\varphi)= \big[Y^i(\varphi)- \frac{4}{a^{3/2}} (\grad_\cG \Lambda)(\varphi)^i\big]\frac{\pd}{\pd \varphi^i}~~,
\ee
where $\Lambda\in \cC^\infty(\cM,\R)$ and $Y\in \cX(\cM)$
satisfy the {\em characteristic system} of the scalar triple
$(\cM,\cG,V)$:
\vspace{-3mm}
{\bf \beqa
\label{S}
&& \Hess_\cG(\Lambda)=\frac{3}{8}\cG \Lambda~~~,~~\cK_\cG(Y)=0~~\\
&& \langle \dd V, \dd \Lambda \rangle_\cG= \frac{3}{4} V \Lambda~~,~~Y(V)=0\nn~~,
\eeqa}
which can also be written in the index-full form: 
\vspace{-2mm}
\beqa
\label{Sindex}
&& \left(\partial_i \partial_j -\Gamma_{ij}^k \partial_k\right) \Lambda=\frac{3}{8}\cG_{ij}\Lambda~~,~~\nabla_i Y_j+\nabla_j Y_i=0~~\\
&& \cG^{ij} \partial_i V \partial_j\Lambda =\frac{3}{4}V \Lambda~~~~~~~~~~~,~~Y^i \partial_i V=0~~~.\nn
\eeqa
\end{thm}
\noindent Here $\Hess_\cG(\Lambda)\eqdef \nabla \dd \Lambda$ (the Hesse tensor of $\Lambda$) and
$\cK_\cG(Y)\eqdef \Sym^2(\nabla Y^\flat)$ (the Killing tensor of $Y$)~.

\subsection{Visible and Hessian symmetries}

\noindent The characteristic system separates into the {\bf $\Lambda$-system}:
\ben
\label{LambdaSys}
 \left(\partial_i \partial_j -\Gamma_{ij}^k \partial_k\right) \Lambda=G_{ij}\Lambda~~,~~G^{ij} \partial_i V \partial_j\Lambda =2V \Lambda~~~\een
and the {\bf $Y$-system}:
\ben
\label{YSys}
 \nabla_i Y_j+\nabla_j Y_i=0~~,~~ Y^i \partial_i V=0~~,
\een
where we use the {\em rescaled scalar manifold metric} $G\eqdef \frac{3}{8}\cG$.
By Theorem \ref{thm:Noether}, strong infinitesimal Noether symmetries have the form:
\ben
\label{Xdec}
X=X_\Lambda+Y~~,
\een
where:
\ben
\label{XLambda}
X_\Lambda\eqdef \frac{\Lambda}{\sqrt{a}}\pd_a- \frac{4}{a^{3/2}} \grad_\cG \Lambda~~,
\een
with $\Lambda\in \cC^\infty(\cM,\R)$ and $Y\in \cX(\cM)$.

\begin{definition}

\
  
\begin{itemize} 
\item A non-trivial vector field $X_\Lambda$ depending on a given
  solution $\Lambda$ of the $\Lambda$-system is called a {\bf Hessian
    symmetry} of $(\cM,\cG,V)$. We say that the scalar cosmological
  model is {\em Hessian} if it admits a non-trivial Hessian symmetry.
\item A non-trivial solution $Y$ of the $Y$-system is called a {\bf
  visible symmetry} of $(\cM,\cG,V)$. We say that the scalar
  cosmological model is {\em visibly symmetric} if it admits a
  non-trivial visible symmetry.
\end{itemize}
\end{definition}
\noindent Let $N(\cM,\cG,V)$ denote the vector space of solutions to the
characteristic system and $N_H(\cM,\cG,V)$ and $N_V(\cM,\cG,V)$ denote
the vector spaces of solutions of the $\Lambda$- and $Y$-systems respectively.

\begin{prop}
There exists a linear isomorphism:
\be
\label{cNiso}
N(\cM,\cG,V) \simeq_\R N_H(\cM,\cG,V)\oplus N_V(\cM,\cG,V)~~.
\ee
In particular, a scalar triple $(\cM,\cG,V)$ admits strong Noether
symmetries iff it is Hessian or visibly symmetric (or both).
\end{prop}

\noindent Since existence of visible symmetries is a well-studied problem, we focus on characterizing
Hessian symmetries.  

\begin{remark}
As shown in \cite{ABL2}, the integral of motion of a Hessian symmetry allows one to
simplify the computation of the number of e-folds along cosmological
trajectories, for which it gives the non-integral formula:
\ben
\label{cNLambda}
N_{t_0}(t)=\frac{2}{3} \log\left[\frac{\Lambda (\varphi(t_0))+
\left(\frac{3}{2}H (t_0)\Lambda(\varphi(t_0))+(\dd\Lambda)(\varphi(t_0))
(\dot{\varphi}(t_0) \right) (t-t_0)}{\Lambda(\varphi(t))}\right]~~
\een
in terms of a solution $\varphi(t)$ of the reduced cosmological equation \eqref{eomsingle}.
\end{remark}

\subsection{Hesse functions and Hesse manifolds}

\begin{definition}
A {\bf Hesse function} of $(\cM,G)$  is a
smooth global solution $\Lambda\in \cC^\infty(\cM,\R)$ of the {\bf Hesse equation}:
\be
\label{HessCond}
\Hess_G(\Lambda)=G \Lambda\,  \Longleftrightarrow \left(\partial_i \partial_j -\Gamma_{ij}^k \partial_k\right) \Lambda=G_{ij}\Lambda~~.
\ee
We denote by $\cS(\cM,G)$ the linear space of Hesse functions of
$(\cM,G)$. When this space is non-trivial, we say that $(\cM,G)$ is a
{\bf Hesse manifold}. (Notice that a Hesse manifold need {\em not} be a
Hessian manifold, i.e. its metric need not be given locally by the
Hessian of a function!)
\end{definition}

\begin{remark}
One can show that any Hesse manifold is non-compact and that $\dim
\cS(\cM,G)\leq n+1$, where $n=\dim \cM$. The space $\cS(\cM,G)$ is
endowed with the {\em Hesse pairing} $(~,~)_G$, a (possibly
degenerate) natural bilinear symmetric pairing which is invariant
under the action of the isometry group of $(\cM,G)$ on Hesse
functions.  Moreover, one can show that $\dim\cS(\cM,G)=n+1$ iff
$(\cM,G)$ is an elementary hyperbolic space form. See \cite{Hesse} for
these and other results on Hesse manifolds.
\end{remark}

\noindent {\bf Example}
The $n$-dimensional Poincar\'e ball $\mD^n\eqdef
(\rD^n,G)$ (with Poincar\'e metric $G$) has $\dim\cS(\cM,G)=n+1$.  Let
$\R^{1,n}=(\R^{n+1},(~,~))$ denote the $(n+1)$-dimensional Minkowski
space, where $(~,~)$ is the canonical Minkowski pairing and let
$E_0,E_1,\ldots, E_n$ be the canonical orthonormal basis of
$\R^{1,n}$. Let $\Xi=(\Xi^0,\Xi^1,\ldots, \Xi^n):\rD^n\rightarrow
\R^{n+1}$ denote the Weierstrass map, where $\Xi^0,\ldots, \Xi^n$ are
the Weierstrass coordinates.
\begin{thm}{\rm \cite{Hesse}}
\label{thm:Hesse}
There exists a bijective isometry $\bLambda:
\R^{1,n}\stackrel{\sim}{\rightarrow} (\cS(\mD^n),(~,~)_{G})$ such
that:
\be
\Lambda_E(u)\eqdef \bLambda(E_\mu)(u)=(E_\mu,\Xi(u))~~,~~\forall u\in \rD^n~~,~~\forall \mu\in \{0,\ldots, n\}~~.
\ee
\end{thm}

\section{Noether symmetries of two-field cosmological models}

\noindent Let $\cM:=\Sigma$ be a connected and oriented surface
($\dim\Sigma=2$), endowed with the complete metric $\cG$. Let $G\eqdef
\frac{3}{8}\cG$ and $V\in\cC^\infty(\Sigma,\R)$. Then: 

\begin{itemize}
\item If the two-field model defined by the scalar triple
  $(\Sigma,\cG,V)$ admits a non-trivial Hessian symmetry, then
  $\Sigma$ is oriented-diffeomorphic to a disk $\rD$, a punctured disk
  $\dot\rD$ or an annulus $A(R)$ of modulus $\mu=2\log R>0$ and $G$ is
  a complete metric of Gaussian curvature $K(G)=-1$ (see
  \cite{ABL2,Hesse,nscalar}). In particular, any such model is a
  generalized two-field $\alpha$-attractor \cite{genalpha} of
  elementary type \cite{elem}, with fixed Gaussian curvature 
  $K(\cG)=-3/8$ of the scalar manifold metric $\cG$.
\item For each of the three cases above, reference \cite{ABL2} gives a
  complete description of the space $\cS(\Sigma,G)$ of Hesse functions
  of $(\Sigma,G)$ and the explicit general form of the scalar potential $V$ which is
  compatible with existence of a non-trivial Hessian symmetry
  generated by a given Hesse function $\Lambda\in \cS(\Sigma,G)$. It
  further gives the special forms of such potentials for which the model
  also admits visible symmetries and describes the space of such
  symmetries in each case.
\end{itemize}

\subsection{Example: Classification of Hessian two-field models whose rescaled scalar manifold is the Poincar\'e disk}

\noindent The following results are proved in \cite[Section 5]{ABL2},
to which we refer the reader for further details and more explicit
formulas written in adapted coordinates on the scalar manifold.  By
Theorem \ref{thm:Hesse}, the general Hesse function on the Poincar\'e
disk $\mD=(\rD,G)$ has the form:
\ben
\label{LambdaB}
\Lambda_B(u)\!=\!(B, \Xi(u))\!=\!B_\mu \Xi^\mu(u)
\!=\!B^0\frac{1+|u|^2}{1-|u|^2}\! -\!
2B^1 \frac{ \Re u}{1-|u|^2} \!-\! 2B^2 \frac{ \Im u}{1-|u|^2}~~(u\in \rD\in \C^2)~~,
\een
where $B=(B^0,B^1,B^2)\in \R^3$ is a non-vanishing 3-vector parameter,
$(~,~)$ is the Minkowski pairing of signature $(1,2)$ on $\R^3$ and
$\Xi=(\Xi^0,\Xi^1,\Xi^2):\rD\rightarrow \R^3$ is the Weierstrass map:
\be
\label{Xi}
\Xi(u)\eqdef \left(\frac{1+|u|^2}{1-|u|^2}, \frac{2 \Re u}{1-|u|^2}, \frac{2 \Im u}{1-|u|^2}\right)~~.
\ee
The following statements hold for the two-field
cosmological model whose scalar manifold is the Poincar\'e disk
$\mD=(\rD,G)$.

\begin{itemize}
\item {\bf When $B\neq 0$ is timelike},
  i.e. $(B,B)=-(B^0)^2+(B^1)^2+(B^2)^2<0$, the two-field model admits
  the Hessian symmetry generated by \eqref{LambdaB} iff the scalar
  potential has the form:
\ben
\label{VTimelike2}
V_B(u)=\omega(n_B(u))\left[\frac{\Lambda_B(u)^2}{(B,B)}-1\right]~~,
\een
where $\omega$ is an arbitrary smooth function
defined on the unit circle and $n_B(u)$ is the 3-vector given by:
\ben
n_B(u)\!=\!\frac{(B,B)\Xi(u)\!-\!(B,\Xi(u)) B}{\sqrt{(B,B)(B,\Xi(u))^2\!
-\!(B,B)^2}}\!=\!\frac{(B,B)\Xi(u)-B\Lambda_B(u)}{\sqrt{(B,B)\Lambda_B(u)^2\!
-\!(B,B)^2}}~.
\een
\end{itemize}

\begin{itemize}
 \item {\bf When $B\neq 0$ is spacelike}, i.e. $(B,B)>0$, the two-field model with scalar manifold
  $\mD$ admits the Hessian symmetry generated by \eqref{LambdaB}
  iff its scalar potential $V$ has the form:
\ben
\label{VSpacelike2}
V_B(u)=\omega(n_B(u))\left[\frac{\Lambda_B(u)^2}{|(B,B)|}+1\right]~~,
\een
where $\omega\in \cC^\infty(\R)$ is an arbitrary smooth function
defined on the real line and $n_B(u)$ is the unit timelike 3-vector
given by:
\ben
\!\!n_B(u)\!\!=\!\!\frac{|(B,B)|\Xi(u)+(B,\Xi(u)) B}{\sqrt{(B,B)^2\!
+\!|(B,B)|(B,\Xi(u))^2}}\!\!=\!\!
\frac{|(B,B)|\Xi(u)+\Lambda_B(u) B}{\sqrt{(B,B)^2\!+\!|(B,B)|\Lambda_B(u)^2}}.
\een
\end{itemize}

\begin{itemize}
\item {\bf When $B\neq 0$ is lightlike}, i.e. $(B,B)=0$, the two-field model with scalar
  manifold $\mD$ admits the Hessian symmetry generated by
  \eqref{LambdaB} iff its scalar potential $V$ has the form:
\ben
\label{VLightlike2}
V_B(u)=\omega(B_0 n_B(u))\frac{\Lambda_B(u)^2}{B_0^2}~~,
\een
where $\omega\in\cC^\infty(\R)$ is an arbitrary smooth function defined
on the real line and $n_B(u)$ is the lightlike 3-vector given by:
\ben
n_B(u)\!=\!\frac{2(B,\Xi(u))\Xi(u)-B}{2(B,\Xi(u))^2}
\!=\!\frac{2\Lambda_B(u)\Xi(u)-B}{2\Lambda_B(u)^2}~.
\een
\end{itemize}

\begin{acknowledgments}
E.M.B. wishes to acknowledge support from PN 18090101/2019, L.A. has received partial support from the Bulgarian NSF grant DN 08/3, while the work of C.I.L. was supported
by grant IBS-R003-D1.
\end{acknowledgments}


\end{document}